# Heterogeneously Integrated Balanced Photodetector on an Ultra-Low Loss Silicon Nitride Delay Line Interferometer

RAHUL CHAWLANI[1,†], FATEMEHSADAT TABATABAEI[2,†], MARK W. HARRINGTON[1], KAIKAI LIU[1], STEVEN M. ZHU[1], JIAWEI WANG[1], MEITING SONG[1], XIANGWEN GUO[2], ANDREAS BELING[2], AND DANIEL J. BLUMENTHAL[1,*]

[1]*University of California at Santa Barbara, Department of Electrical and Computer Engineering, Santa Barbara, CA, USA*
[2]*University of Virginia, Department of Electrical and Computer Engineering, Charlottesville, Virginia 22904, USA.*
[†]*These authors contributed equally.*
**danb@ucsb.edu*

**Abstract:** Thin core silicon nitride photonics enables ultra-low loss, CMOS foundry compatible integration that supports wavelengths from the visible to shortwave infrared. Applications that can benefit from the resulting lower cost, improved robustness, and portability include quantum sensing and computing, ultra-low noise microwave generation, optical clocks, optical gyros, coherent fiber communications, and fiber sensing. An important next step is integration of functional circuits and systems on chip with heterogeneous integration of active components such as high-performance photodetection. Yet to date integrated high-performance photodetectors on the thin film silicon nitride platform has remained elusive. In this work, we demonstrate heterogeneous integration of an InGaAs on InP substrate Modified Uni-Traveling Carrier balanced photodetector with a 15-meter-long unbalanced thin core silicon nitride Mach-Zehnder Interferometer with a bandwidth of 0.92 GHz and a responsivity of 0.305 A/W at 1550 nm with a propagation loss as low as 2.5 dB/m at 1600 nm. Using this circuit we demonstrate two functions, a meter-scale differential interferometer laser stabilization circuit achieving a nearly 23 dB noise suppression at 1 kHz offset and an optical frequency discriminator frequency noise measurement with high sensitivity across 6 orders of magnitude from 10 Hz to 10 MHz. These results demonstrate that the high performance of thin core silicon nitride devices can be combined with integrated high-performance photodetection to realize on-chip stabilized lasers and circuits and pave the path towards full systems on chip.

## 1. Introduction

Low noise, stabilized lasers are a prerequisite for applications such as quantum computing [1,2], optical clocks [3–5], quantum sensing [6,7], low-noise microwave generation [8–11], fiber sensing [12], and coherent communications [13–15]. Typical methods such as Pound-Drever-Hall (PDH) locking [16–18] and interferometric based methods [19–22] require a sharp discriminator signal from an optical reference cavity, photodetector, and locking electronics to provide feedback to the laser. Full integration of this system is a necessary step for the applications listed above to reduce size, weight, and power (SWaP) [23,24] while ensuring portability [25] and scalability [26,27].

The silicon nitride ($Si_3N_4$) platform with thin, high-aspect ratio waveguides [28] is a leading candidate for full scale integration of stabilized lasers due to its wide transparency window and low propagation loss from visible to shortwave infrared (SWIR) [29–33], CMOS-compatibility [34,35], ultra-low phase noise and low linewidth lasers [27,34,36–41], and quantum and atomic systems on chip [42,43]. Such low-loss technology has the ability to unlock meter-scale devices such as large mode volume coil resonators with a low thermo-refractive noise (TRN) floor to reduce integral linewidth and frequency noise of a stabilized laser [9,44–46]. However, stabilization of a laser to an on-chip reference cavity via PDH locking requires bulk modulators or frequency shifters [17]. While integrated stress-optic modulators with lead zirconate titanate (PZT) [47,48] and aluminum nitride (AlN) [49], electro-optic modulators via thin-film lithium niobate [50], and acousto-optic modulators (AOM) [51,52] have been demonstrated on chip, they still increase the complexity and power consumption of the locking setup – limiting applications requiring hundreds of stabilized lasers. Integrated unbalanced Mach-Zehnder Interferometers (MZI) with meter-scale delay arms have been demonstrated on the silicon nitride platform as an optical frequency discriminator for modulator-free laser stabilization [53] and self-homodyne frequency noise measurements [54]. Furthermore, detection of the error signal from both output ports of the MZI with a balanced photodetector (BPD) doubles the discriminator slope and suppresses laser intensity noise [53,55].

However, the lack of integrated balanced photodetectors on this low-loss platform represents a bottleneck to integrability and scalability of these systems. Heterogeneously integrated photodetectors on silicon nitride with the necessary responsivity and bandwidth to provide electronic feedback for laser stabilization have had limited demonstration, having only been shown on thick waveguide cores at C band [56–61] and O band [62–64]. These thicker waveguide cores confine the light more effectively, increasing propagation loss in comparison to thin, high-aspect ratio waveguides [28]. Furthermore, other methods have involved complex waveguide to photodetector coupling schemes that are difficult to fabricate [65,66].

In this work, we demonstrate a heterogeneously integrated balanced photodetector on a high-aspect ratio, low-loss silicon nitride platform. This BPD has a total internal responsivity of 0.305 A/W and a bandwidth of 0.92 GHz suitable for locking applications while the silicon nitride platform has a propagation loss of 2.5 dB/m at 1600 nm. We demonstrate this technology with an on-chip 15 meter long, unbalanced MZI acting as an optical frequency discriminator (OFD). Using this device, we stabilize a laser and achieve 23 dB frequency noise suppression compared to an unlocked laser at 1 kHz frequency offset. Furthermore, we demonstrate frequency noise characterization via the delayed self-homodyne technique. We present a comparison of this work's responsivity, propagation loss, and waveguide core thickness to other works in Table 1.

**Table 1. Comparison of Directly Integrated InGaAs Photodetector Technology on $Si_3N_4$ at 1550 nm**

| Ref. | $Si_3N_4$ Thickness (nm) | Photodetector Type | Responsivity (A/W) | $Si_3N_4$ Propagation Loss (dB/m) |
|---|---|---|---|---|
| **This Work** | **80** | **MUTC BPD** | **0.305** | **3.1** |
| [56] | 400 | MUTC | 0.94 | - |
| [57] | 300 | UTC | 0.3 | <100 |
| [58] | 400 | MUTC BPD | 0.72 | - |
| [59] | 800 | MUTC | 0.76 | 15.8* |
| [60] | 400 | P-I-N PD | 0.68 | 600 |

***BPD**: Balanced photodetector. **UTC**: Uni-traveling carrier. **MUTC**: Modified uni-traveling carrier. **$Si_3N_4$**: Silicon Nitride. *Calculated from the given resonator Q values.*

## 2. Device Overview and Fabrication

The device is a 15-meter MZI on a low-loss $Si_3N_4$/Si platform whose output waveguides couple to a heterogeneously integrated BPD on InP substrate (Fig. 1a). The input waveguides (Channels 1 and 2) interfere with each other in the waveguide with one arm having a 15-meter delay, and the interference is detected on the output BPD (Channels 3 and 4) whose difference provides the balanced signal (Fig. 1b). An image of the input waveguides and pads for probing are provided (Fig. 1c).

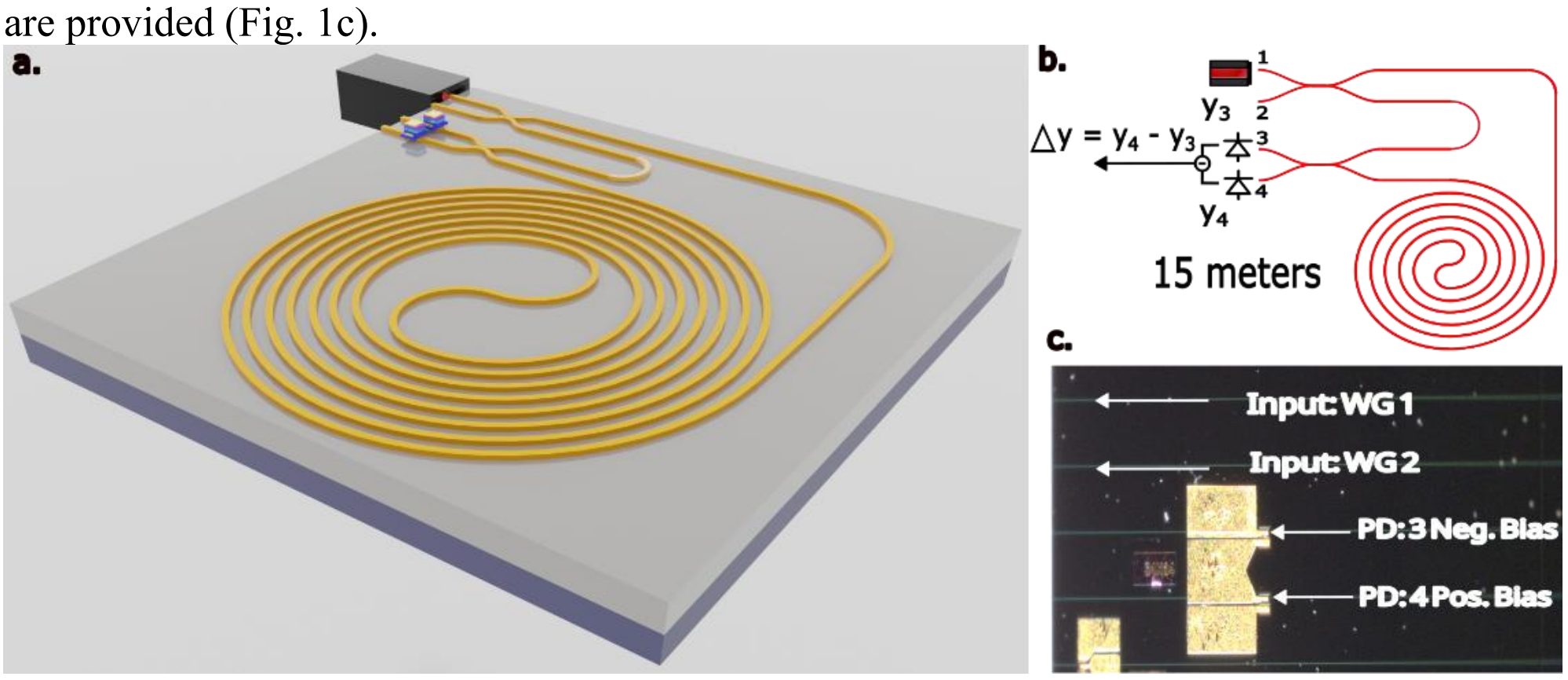

***Fig. 1. Overview of Heterogeneously Integrated BPD and MZI**. (a) 3D model of full device with off-chip laser (b) Schematic of device operating principle. (c)Photograph of integrated BPD and input waveguides with corresponding channel numbering. Image was digitally enhanced for clarity. **BPD**: Balanced Photodetector. **MZI**: Mach-Zehnder Interferometer. **WG**: Waveguide.*

A complete fabrication flow chart of the photodetector to chip bonding can be seen in Fig. 2a. The $Si_3N_4$ photonic passive device is a high-aspect ratio waveguide with 80 nm thick x 6 microns wide core design using LPCVD nitride with 15 µm thermal oxide bottom cladding and 6 micron TEOS-PECVD upper oxide cladding. (see Supplemental Information Section I for further details). The input waveguides have a taper width of 1.5 µm to increase fiber to chip coupling. Two directional couplers are 1.4 mm long and designed to support 50:50 coupling. Furthermore, the output waveguide to the photodetectors also has 1.5 µm tapers. Finally, the device is designed to support only the TE0 mode.

A layer stackup of the individual photodetectors can be seen in Fig. 2b. After the passive device is fabricated as normal, the bond window is etched such that the upper cladding in the area of interest is etched down all the way (6 microns). The photodetector is an InGaAs on InP substrate MUTC (Modified Uni-Traveling Carrier) designed for high quantum efficiency. The unprocessed epitaxial stack was bonded to the photonic device using SU-8 as the adhesive layer. p- and n- doped mesas are formed via wet etching. Gold metal is then deposited and lifted off. The two PDs are balanced via fabrication of a metal air-bridge plating between the two photodetectors. The silicon dioxide upper cladding is then redeposited over the whole chip with openings etched on the metal pads for probing the BPD. Contact with the photodetectors is accomplished with a custom RF probe. Images of this process can be found in Supplemental Fig. 1.

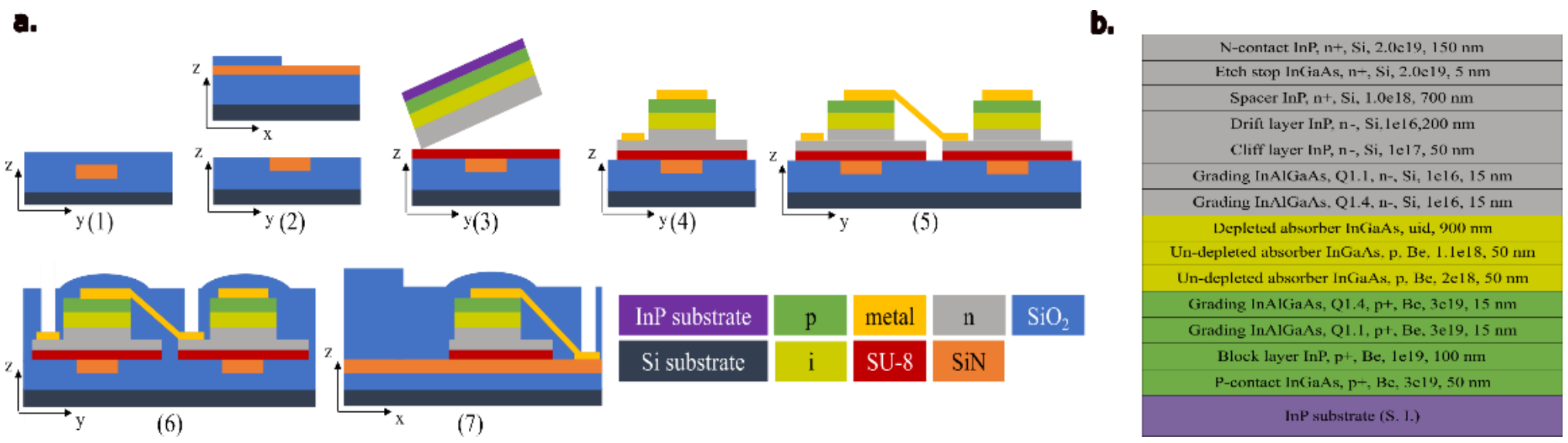

***Fig. 2. Fabrication details of heterogeneously integrated BPD and MZI** (a) Fabrication process steps: (1) Fabricated low-loss $Si_3N_4$ chip. (2) Etching the bonding window. (3)*

*Bonding using SU-8 as the adhesive layer. (4) Metal deposition/lift-off. Forming p- and n-mesa via wet etch. (5) Air-bridge plating for balanced photodetectors. (6) Front view after $SiO_2$ redeposition and opening (7) Side view after $SiO_2$ redeposition and opening (b) Layer stackup of individual InGaAs on InP photodetector with dopants, doping level, and thickness.* ***BPD****: Balanced Photodetector.* ***MZI****: Mach Zehnder Interferometer*

## 3. Results

We measure the C-V characteristics of each individual PD after upper cladding redeposition and note a capacitance of 320 fF at -3 V applied bias (Fig. 3a). We also measure the dark I-V characteristics before and after the silicon dioxide redeposition (Fig. 3b). Before redeposition, the dark current at -3 V bias was 8.6 nA and 6.8 nA with an increase after redeposition to 5.6 µA and 0.48 µA on the PD of channels 3 and 4, respectively. A diagram of the electrical setup can be found in Supplemental Fig. 2a. We believe the increase in dark current seen after $SiO_2$ redeposition can be attributed to surface leakage, and that future work can focus on improvements in passivation before cladding redeposition [67].

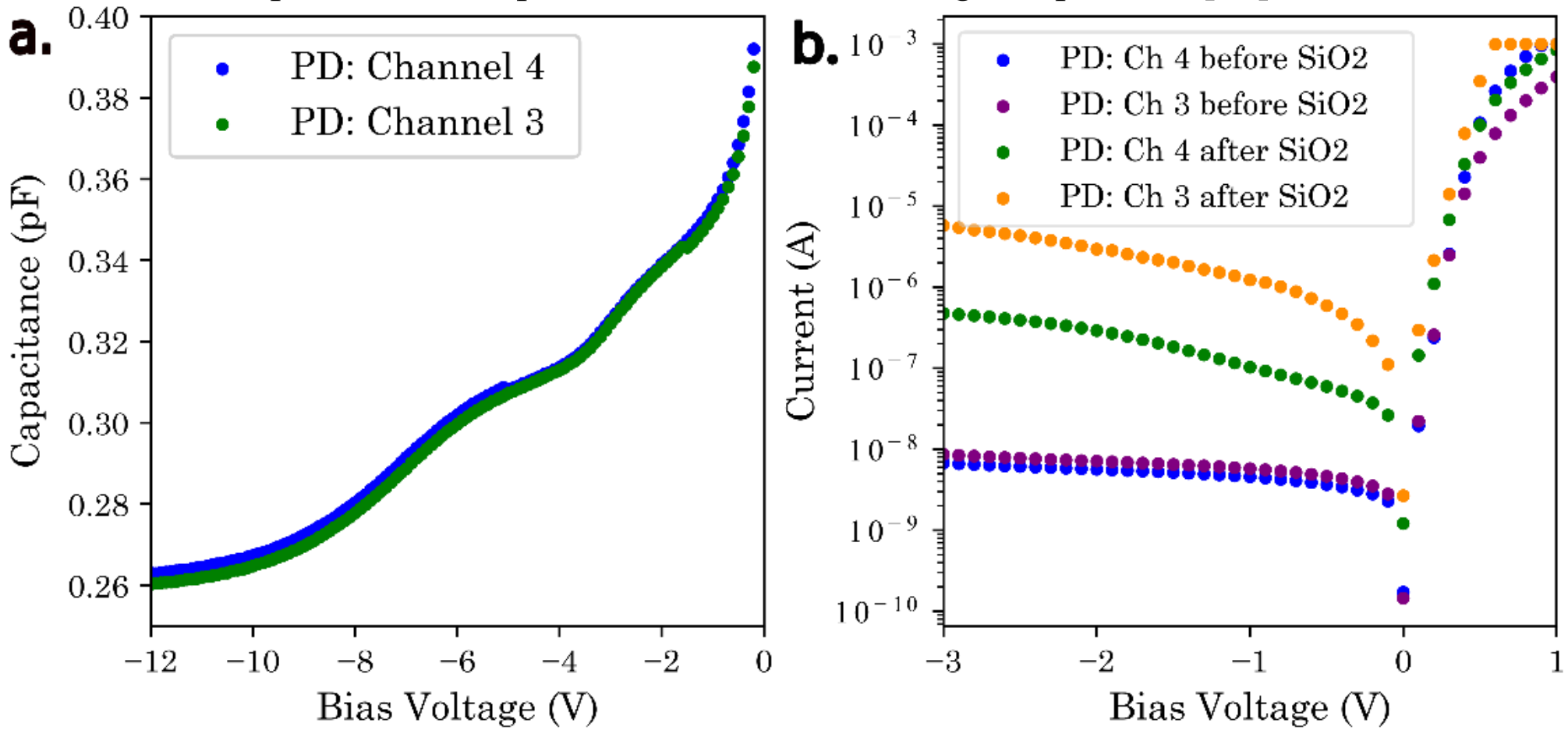

***Fig. 3****.* ***C-V and I-V characteristics of individual photodiodes****. (a) Capacitance-voltage relationship of individual PDs. We measure a capacitance of 320 fF at –3 V bias. (b) Current-voltage relationship before and after oxide redeposition. We see an unequal increase in dark current after oxide redeposition.* ***PD****: Photodiode*

We then measure the characteristics of the 15-meter unbalanced MZI itself using a 200-meter calibrated fiber MZI as a reference, measuring an FSR of 13 MHz at 1550 nm (Fig. 4a). Measurements were done with a widely tunable C band ECDL (External Cavity Diode Laser, Newport Velocity TLB-6730) from 1550 to 1630 nm. The extinction ratio (ER) of the MZI as a function of wavelength can be found with an ER of 0.1 dB at 1550 nm and 0.24 dB at 1600 nm. Fitting the measured MZI signal with a theoretical curve [54,68], the coupling ratio for the two directional couplers was confirmed to be 50:50 and the propagation loss of the waveguide can be extracted (Fig. 4b). We report a propagation loss of about 3.1 dB/m at 1550 nm and 2.5 dB/m at 1600 nm. Before the BPD was bonded, the passive photonic chip had a propagation loss of about 1.5 dB/m at 1550 nm and 1.25 dB/m at 1600. We attribute this higher than normal [31] loss for this waveguide geometry due to a low point resolution on the mask of the 15 m delay arm. A diagram of the electrical setup used to capture this data can be found in Fig. 2b.

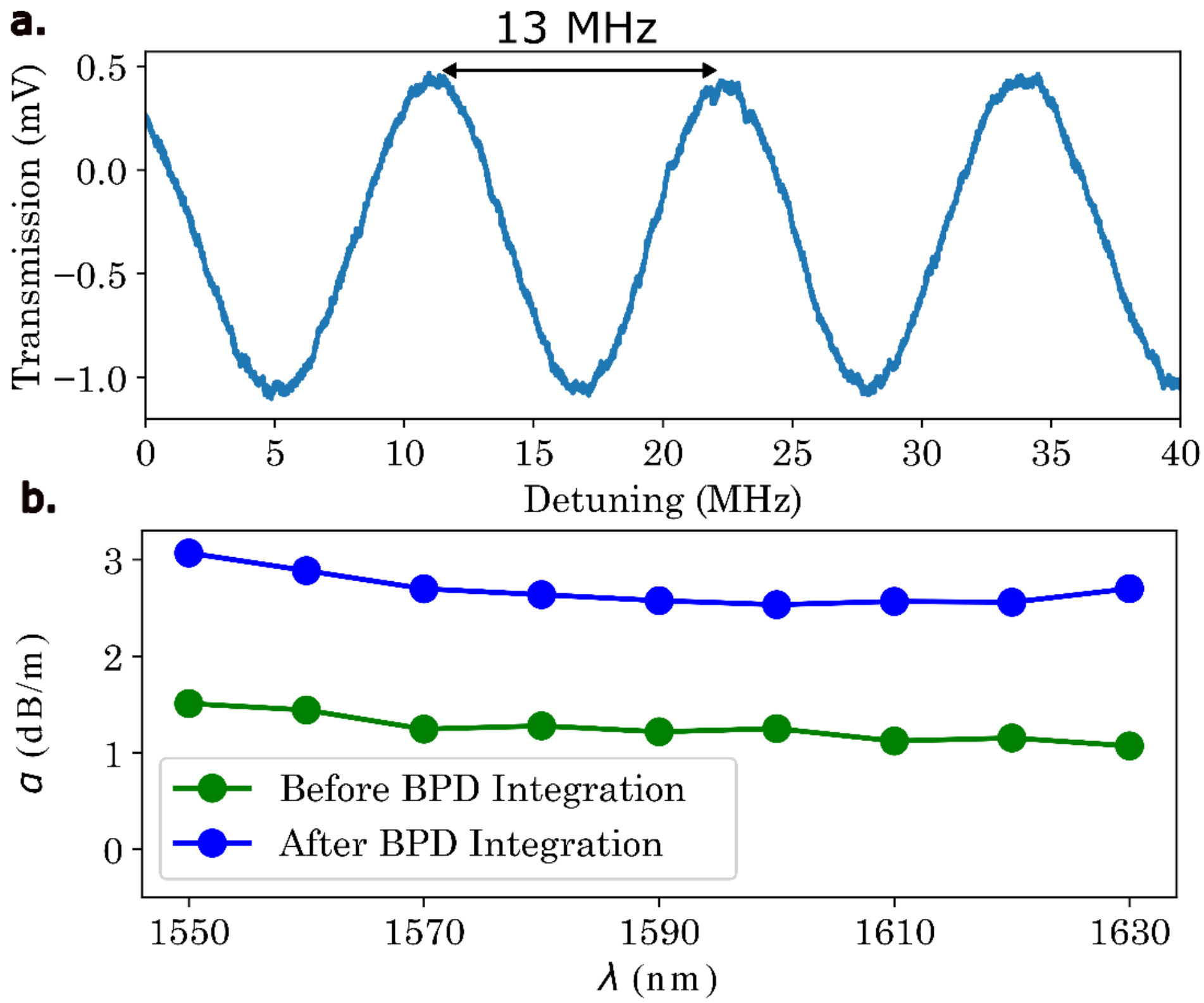

***Fig. 4. MZI characteristics with BPD**. (a) Transmission of the MZI signal swept over a small frequency range. We measure an FSR of 13 MHz. (b) Propagation loss from 1550 to 1630 nm before and after BPD integration. **α**:Propagation loss. **FSR**: Free-spectral range.*

Using the same electrical setup used in Fig. 4, we can also measure the internal responsivity of each PD and the total responsivity of the BPD at 1550 nm at a bias voltage of ±3 V (Fig. 5). We account for both the internal propagation loss of the MZI, and the facet loss (2.725 dB/facet). We measure a total internal responsivity of 0.305 A/W and an internal responsivity of 0.173 A/W and 0.132 A/W for PD channel 3 and 4 respectively, corresponding to a quantum efficiency of 0.139 and 0.106.

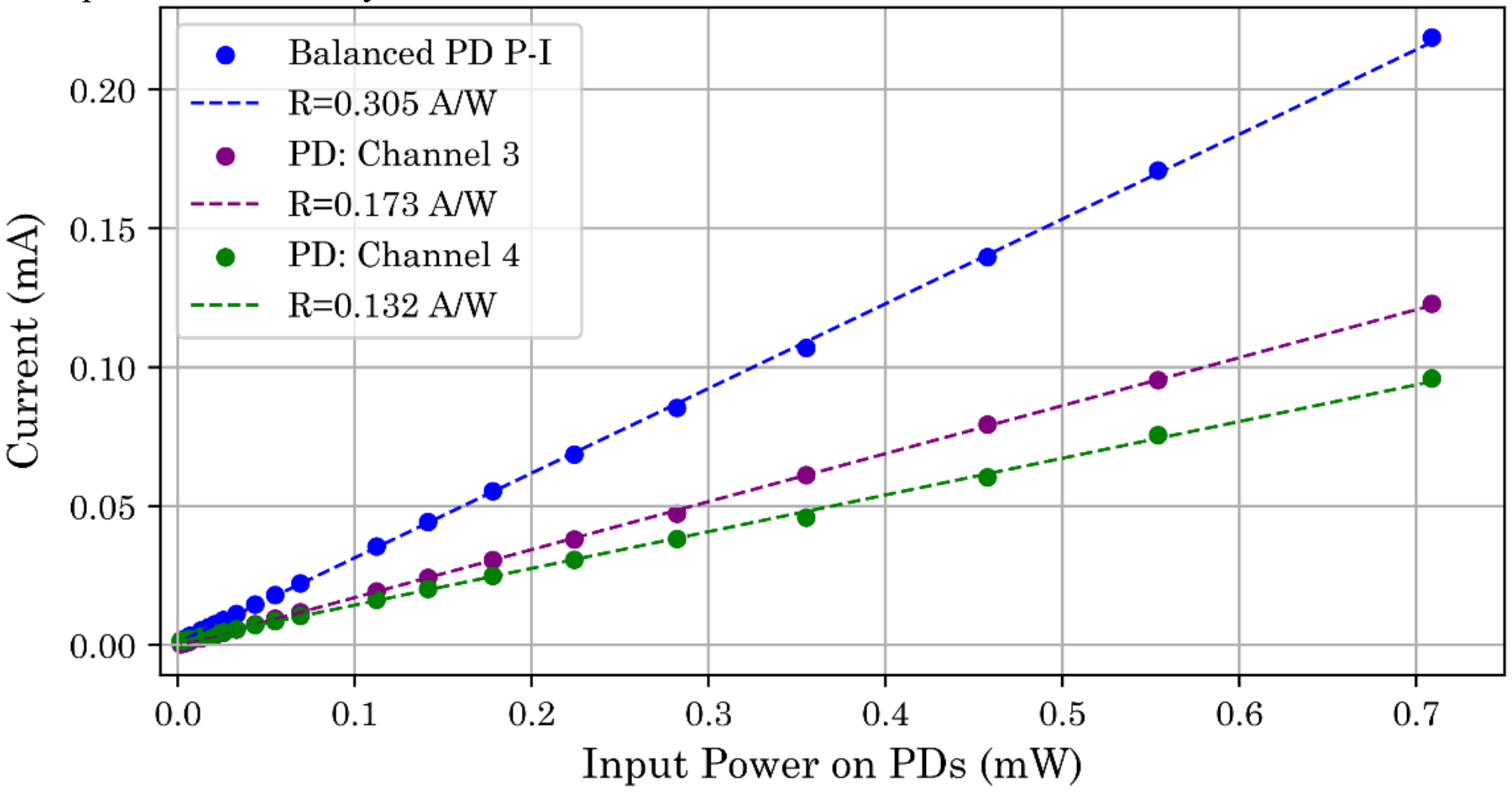

*__Fig. 5. P-I Measurement of heterogeneously integrated BPD at 1550 nm.__ Measured optical power - current relationship. We apply a linear fit to calculate internal responsivity for each PD and total responsivity. We adjust the x-axis by accounting for facet loss and propagation loss. __PD__: Photodetector.*

Bandwidth measurements were done to understand the bandwidth of the BPD and any contributions from the MZI. Light at 1550 nm passed through an EDFA then an intensity modulator biased at quadrature with a modulation frequency swept by a PNA (Performance Network Analyzer, Keysight N5247B PNA-X) before entering the heterogeneously integrated MZI and BPD. The balanced, detected electrical signal is then measured and directly entered into the PNA (Fig. 6a). A slow ramp signal was sent to the ECDL to control the mirror during this measurement to ensure the bandwidth was taken at the MZI quadrature point. The frequency response of the combined heterogeneously integrated BPD and MZI in combination with the TIA can be found in Supplemental Section III.

Both the 3 dB and 6 dB cutoff frequency were measured without the TIA with a 3 dB bandwidth of 0.92 GHz and a 6 dB bandwidth of 1.79 GHz (Fig. 6b). Inserting the TIA, we measure a 3 dB bandwidth of 20.7 MHz and a 6 dB bandwidth of 23.8 MHz (Supplemental Fig. 4). While the TIA restricts the bandwidth, it is still high enough for many laser stabilization applications where the photodetector bandwidth must be higher than the modulation frequency [18] while faster, low-noise TIAs have been reported [69,70]. In this analysis, we neglect the contribution from the 15-meter MZI's amplitude frequency response due to its low ER with a theoretical MZI amplitude frequency response with the contributions of this low ER plotted in Fig. 6b. Further analysis can be found in Supplemental Section IV.

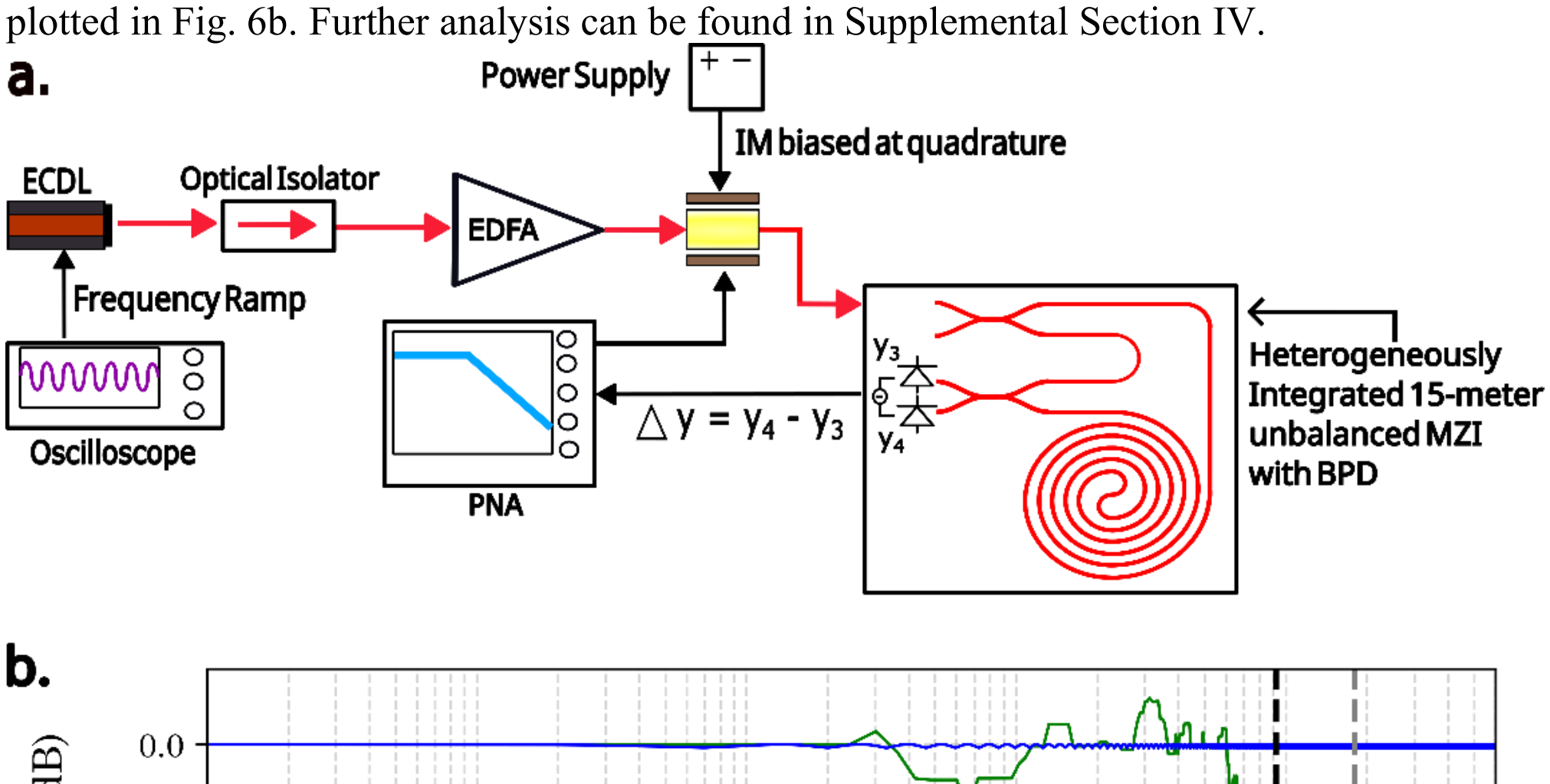

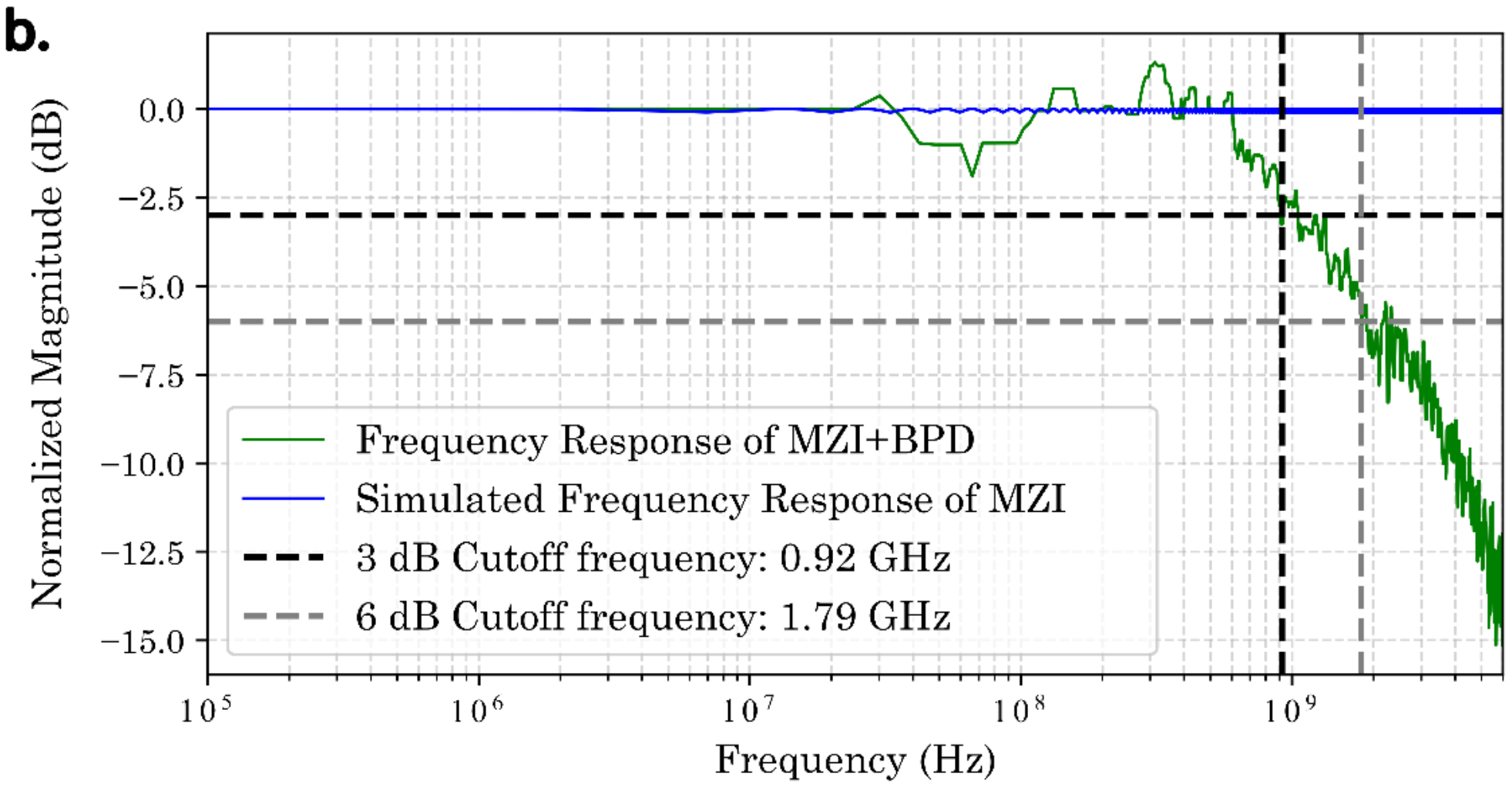

*__Fig. 6. Bandwidth Measurement of the Heterogeneously Integrated BPD and MZI.__ (a) Measurement setup using a PNA to sweep frequencies across an electro-optic modulator biased at quadrature. An EDFA was used to reduce the effects of ripples. (b) Normalized frequency response of the BPD and MZI in green. A 3 dB cutoff frequency of 0.92 GHz was found. Simulated amplitude frequency response of the MZI given the calculated extinction ratio is provided in blue. __ECDL__: External Cavity Diode Laser. __EDFA__: Erbium-doped Fiber Amplifier. __IM__: Intensity Modulator. __PNA__: Performance Network Analyzer. __BPD__: Balanced Photodetector. __MZI__: Mach-Zehnder Interferometer.*

The heterogeneously integrated BPD and MZI were then used for modulator-free laser locking. We note the discriminator slope (defined as $D = \frac{V_{pp}}{FSR}$) of this MZI was measured to be 0.146 mV/MHz at 5 dBm input power without any external electrical amplification and 35.1 mV/MHz at 8.9 dBm input power with the TIA. A diagram of the locking setup can be found in Fig. 7a and a diagram of the electrical setup to access the BPD can be found in Supplemental Fig. 2c. An ECDL at 1550 nm is sent through an isolator, and 98% of the optical signal is tapped to send through the heterogeneously integrated BPD and MZI. The electrical signal is then amplified and sent through a servo (Vescent D2-125-PL) that controls the ECDL current and an auxiliary loop with the ECDL mirror. The laser is locked at the quadrature point of the MZI. The frequency noise of the laser is measured using a 200-meter fiber MZI which is detected by a bulk BPD (Thorlabs PDB450C) with the adjustable gain set to $10^4$. We compare the locked and free-running noise performance in Fig. 7b and measure a 23 dB reduction in frequency noise at 1 kHz offset frequency. ￼￼ for further details).

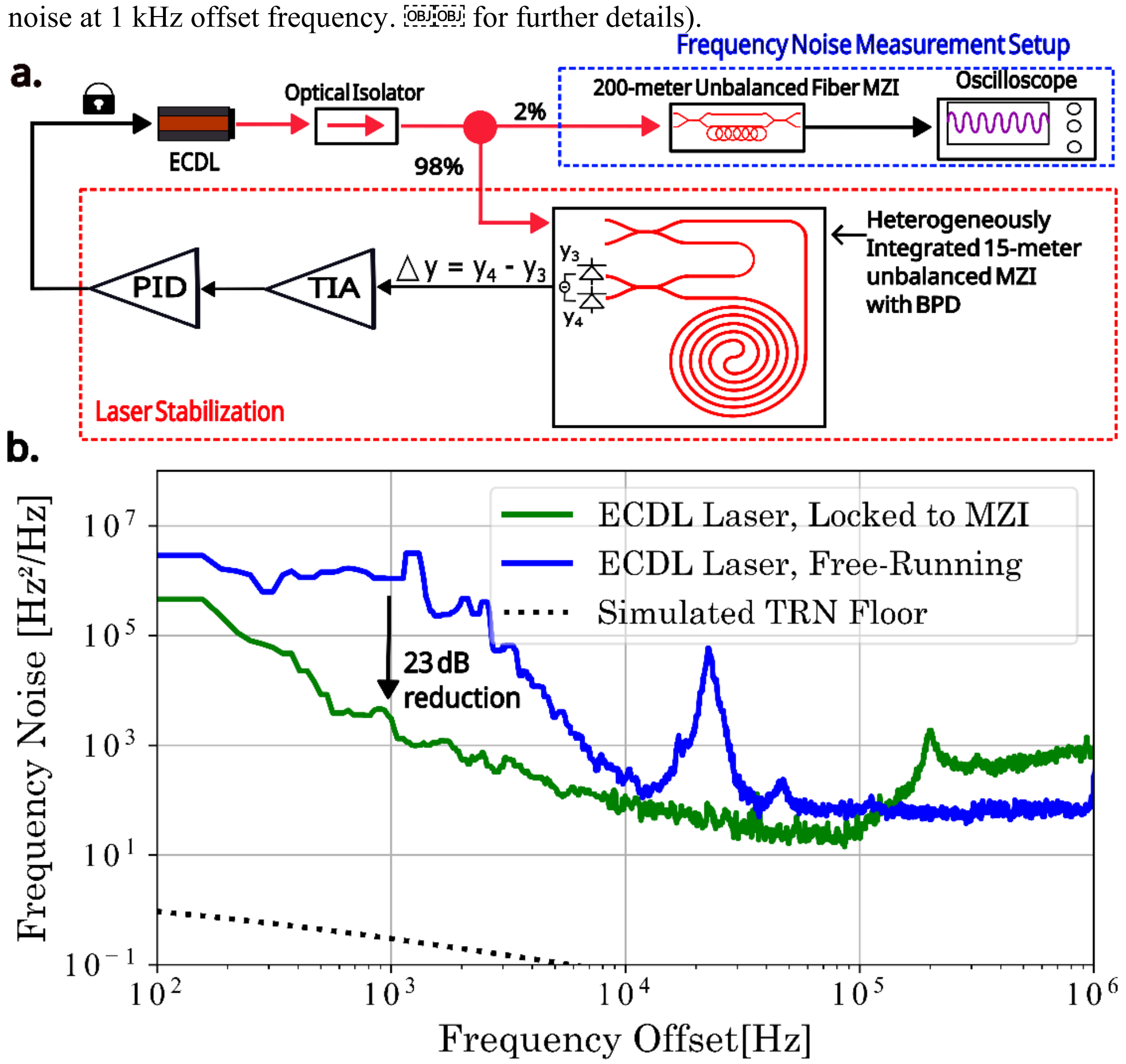

***Fig. 7. Modulation-free laser stabilization with heterogeneously integrated BPD and MZI at 1550 nm**. (a) Laser stabilization setup with fiber frequency noise measurement setup. Components within dashed red box were used for laser stabilization while components within the dashed blue box were used for laser frequency noise measurement. (b) Frequency noise performance of the ECDL when both locked and unlocked to heterogeneously integrated 15-meter MZI and BPD. **ECDL**: External cavity diode laser. **MZI**: Mach-Zehnder Interferometer. **TIA**: Transimpedance amplifier. **BPD**: Balanced photodetector. **TRN**: Thermo-refractive noise.*

We then use the heterogeneously integrated BPD and MZI as an optical frequency discriminator to measure the frequency noise of a free-running ECDL at 1550 nm. A setup can be found in Fig. 8a. The free-running laser is sent through an isolator and then the heterogeneously integrated BPD and MZI. The output electrical signal is amplified and measured on an oscilloscope where the PSD of the signal is taken. The scope is triggered at the quadrature point of the MZI as the laser frequency noise passes through the zero-crossing and zoomed in at the quadrature point to reduce digitization noise. A comparison of the free-running ECDL frequency noise measured via the heterogeneously integrated BPD and MZI and an unbalanced fiber MZI with a calibrated FSR of 18 MHz with an external BPD can be found in Fig. 8b. Finally, we perform a noise floor analysis by measuring the frequency noise contributions of the oscilloscope (Keysight DSOX1204G), integrated BPD, and TIA noting that the TIA and BPD add noise at frequency offsets at about 2 MHz above the measured frequency noise of the 18 MHz fiber MZI. For bandwidth and frequency noise plots (Fig. 6b, 7b, 8b) we use a median averaging filter to smooth out the plots. We used raw data to calculate bandwidth and frequency noise reduction.

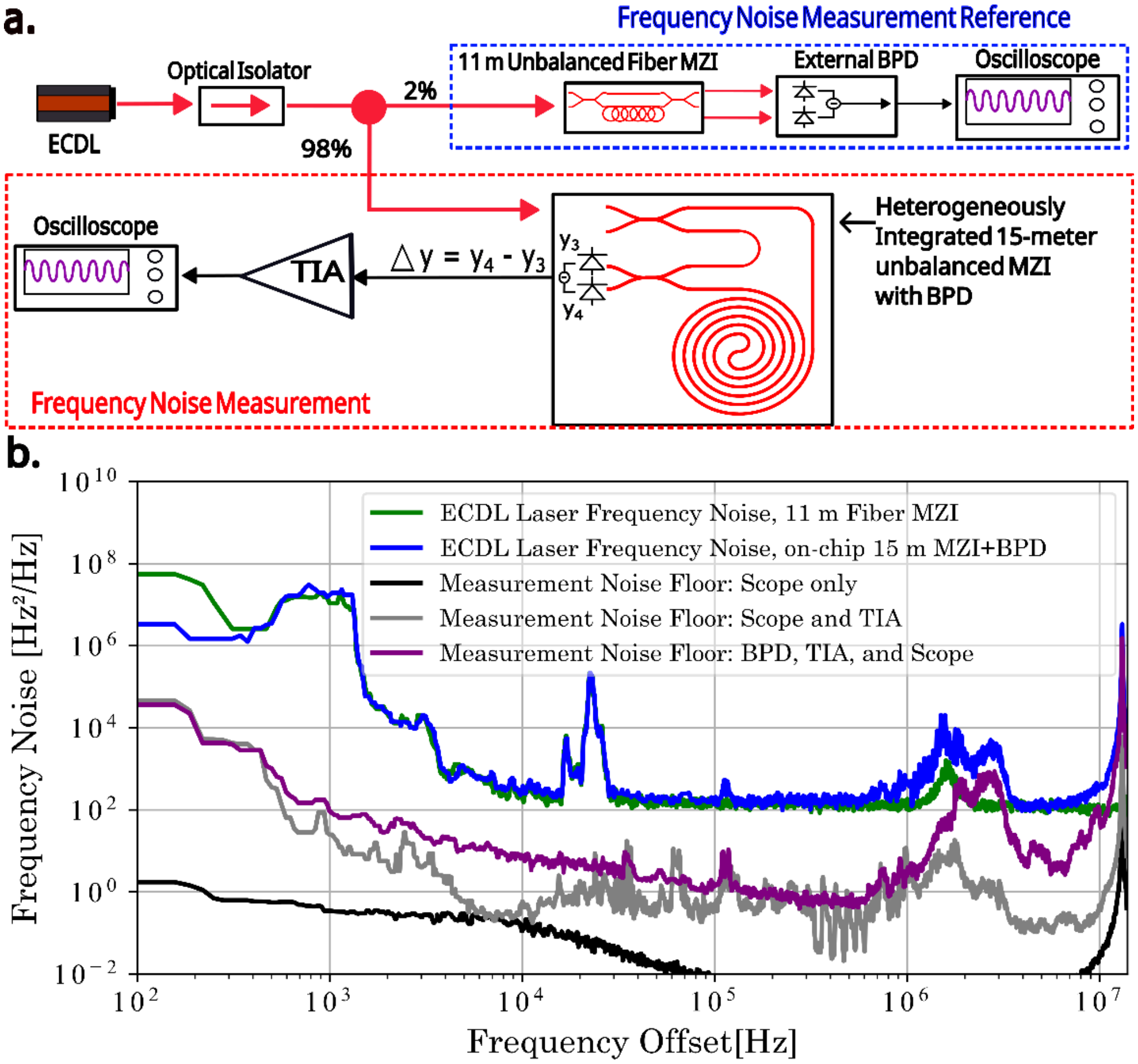

***Fig. 8. Frequency noise measurement with heterogeneously integrated BPD and MZI at 1550 nm**. (a) Laser stabilization setup with fiber frequency noise measurement setup. Components within dashed red box were used as an optical frequency discriminator to measure frequency noise while components within the dashed blue box were used as a reference noise measurement system with an 18 MHz MZI. (b) Comparison of frequency noise measurement between the reference fiber MZI with external, bulk BPD and the heterogeneously integrated MZI and BPD. Noise sources from the oscilloscope, TIA, and BPD are shown. Noise spikes at 13 MHz are due to the integrated MZI transfer function. **ECDL**: External cavity diode laser. **MZI**: Mach-Zehnder Interferometer. **TIA**: Transimpedance amplifier. **BPD**: Balanced photodetector. **Scope**: Oscilloscope.*

## 4. Discussion

We have demonstrated significant advancement in heterogeneously integrated low-loss photonics for laser stabilization and frequency noise measurements with the demonstration of a 15-meter MZI coupled to an integrated BPD. This planar integrated circuit is compatible with existing technology in silicon nitride such as on-chip lasers, stress optic modulators, nonlinear photonics, and other components that are comparable to bulk and fiber optic solutions. We report that the heterogeneously integrated circuit preserves both the low loss characteristics of high-aspect ratio silicon nitride waveguides (2.5 dB/m at 1600 nm) and the high responsivity of the BPD (0.305 A/W at 1550 nm). We note a 3 dB cutoff frequency of the BPD to be 0.92

GHz while use of this circuit in applications demonstrates a 23 dB frequency noise reduction at 1 kHz and agreement with a fiber MZI for use for frequency noise measurement.

In the future, integration of external cavity lasers [36,72,73], ultra-high Q cavities [27,29], and cavity-coupled MZIs [19,74] will be possible for further laser stabilization and system on a chip complexity. Future work can also focus on heterogeneously integrated lasers with metal or stress-optic tuning [47,48,75] and integrated electronics [76] for a completely stabilized chip-scale laser. Finally, with advances in low-loss silicon nitride waveguides and devices at visible wavelengths [77,78], this method can be adapted for different photodetector materials [79–81]. This technology will advance a suite of technologies for precision applications requiring fully integrated photonic circuits.

## 5. Back matter

**Funding.** This work is based in part by funding from DARPA GRYPHON award HR0011-22-2-0008. The views and conclusions contained in this document are those of the authors and should not be interpreted as representing the official policies of DARPA or the U.S. government.

**Acknowledgment.**
We acknowledge Karl Nelson of Honeywell for help fabricating the 15-meter MZI photonic chip. We also acknowledge Andrei Isichenko and David Heim for helpful discussion on this work. We also acknowledge Joshua Hansen for help designing and testing the TIA.

### Disclosures

Dr. Blumenthal has consulted for Infleqtion and owns stock in the company. The other authors declare no potential conflict of interest.

**Data availability.** Data underlying the results presented in this paper are not publicly available at this time but may be obtained from the authors upon reasonable request.

### Supplementary Document

See Supplement 1 for supporting content.

# Heterogeneously Integrated Balanced Photodetector on an Ultra-Low Loss Silicon Nitride Delay Line Interferometer:

# Supplemental document

**Section I. Photonic Chip Fabrication and Bonding Process**

To fabricate the photonic chip, thermal oxide lower cladding is grown atop a 200 mm diameter silicon wafer to 15 µm thick. We deposit a 75 nm thick $Si_3N_4$ layer using low-pressure chemical vapor deposition (LPCVD) on top of the lower oxide cladding. Patterning is then achieved after photoresist spinning with a DUV stepper. We dry etch in an inductively coupled plasma etcher using $CHF_3/CF_4/O_2$ chemistry with a standard Radio Corporation of America (RCA) cleaning process applied after etching. An additional thin $Si_3N_4$ layer is deposited followed by a 1100 ºC anneal for 30 minutes in an oxygen atmosphere. A 6 µm oxide upper cladding is deposited using plasma-enhanced chemical vapor deposition (PECVD) with tetraethoxysilane (TEOS) as a precursor. Finally, high temperature annealing with a first step of 1050 ºC for 7 hours and 1150 ºC for 2 hours is done.

A step-by-step bonding and photodiode fabrication processing flowchart can be seen in Fig. 2. Of critical importance is the bonding window of the original passive photonic chip. Images of this process are seen in Supplemental Fig. 1. A layout of the whole passive photonic chip with the selected bonding window and dicing lines is seen in Supplemental Fig. 1a. After upper oxide removal, the epi stack of the unprocessed photodiodes is bonded. The edges of the epi stack are protected with photoresist to minimize the lateral etching (Supplemental Fig. 1b). We then remove the substrate (Supplemental Fig. 1c). PD fabrication is accomplished as described in the main text, and the chip is then diced to access the waveguides' input (Supplemental Fig. 1d).

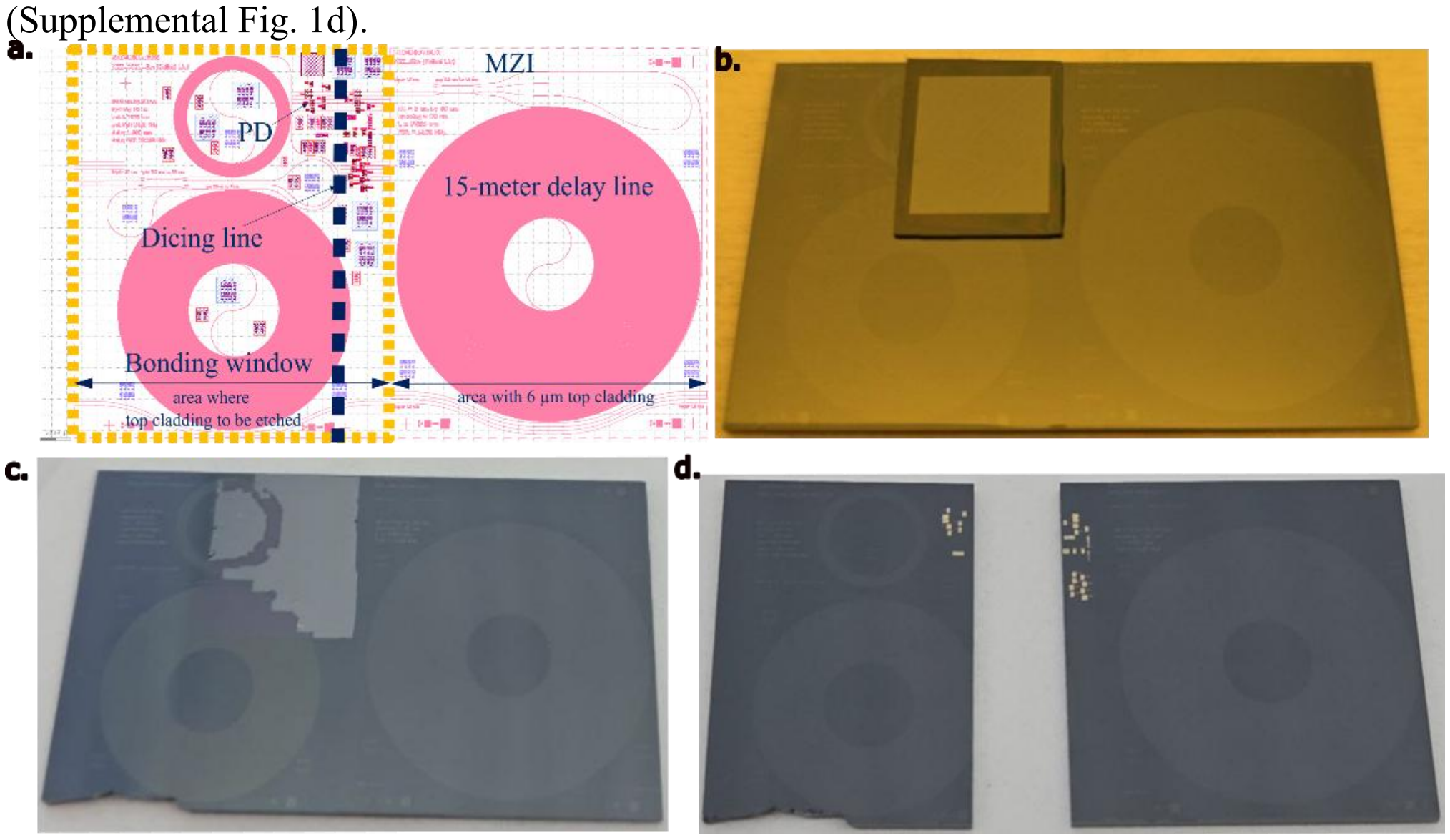

***Supplemental Fig. 1: Details on photodiode die bonding process to photonic chip.*** *(a) Layout of the chip including the PDs and the MZI. (b) After bonding epi. The edges of the epi were protected by photoresist before substrate removal to avoid lateral etching . (c) After substrate removal. (d) After PD fabrication is completed, the chip is diced. The size of the original $Si_3N_4$ chip is 42 µm × 25 µm.*

**Section II. Electrical Setup**

For testing and use in applications such as laser stabilization and frequency noise measurement, three different electrical configurations were used to access and power the BPD. All three can be found in Supplemental Fig. 2. The BPD is accessed via a custom RF balanced probe. Both PDs are biased at ±3 V with the positive channel of the power supply connected to the probe while the two grounds are connected to each other and to the DC port of a low-frequency bias tee. The AC port is connected to an RF power meter, oscilloscope, or PNA while the complete signal is connected to the balanced port of the BPD (Supplemental Fig. 2a). In Supplemental Fig. 2b, where individual channels of the PD are wished to be seen to understand loss in Fig. 4 or individual responsivity in Fig. 5, a small resistor is placed in series with the PD and can be probed. In Supplemental Fig. 1c, where the complete signal is needed for laser stabilization, the bias tee is removed, and a TIA is inserted with details in Supplemental Section III.

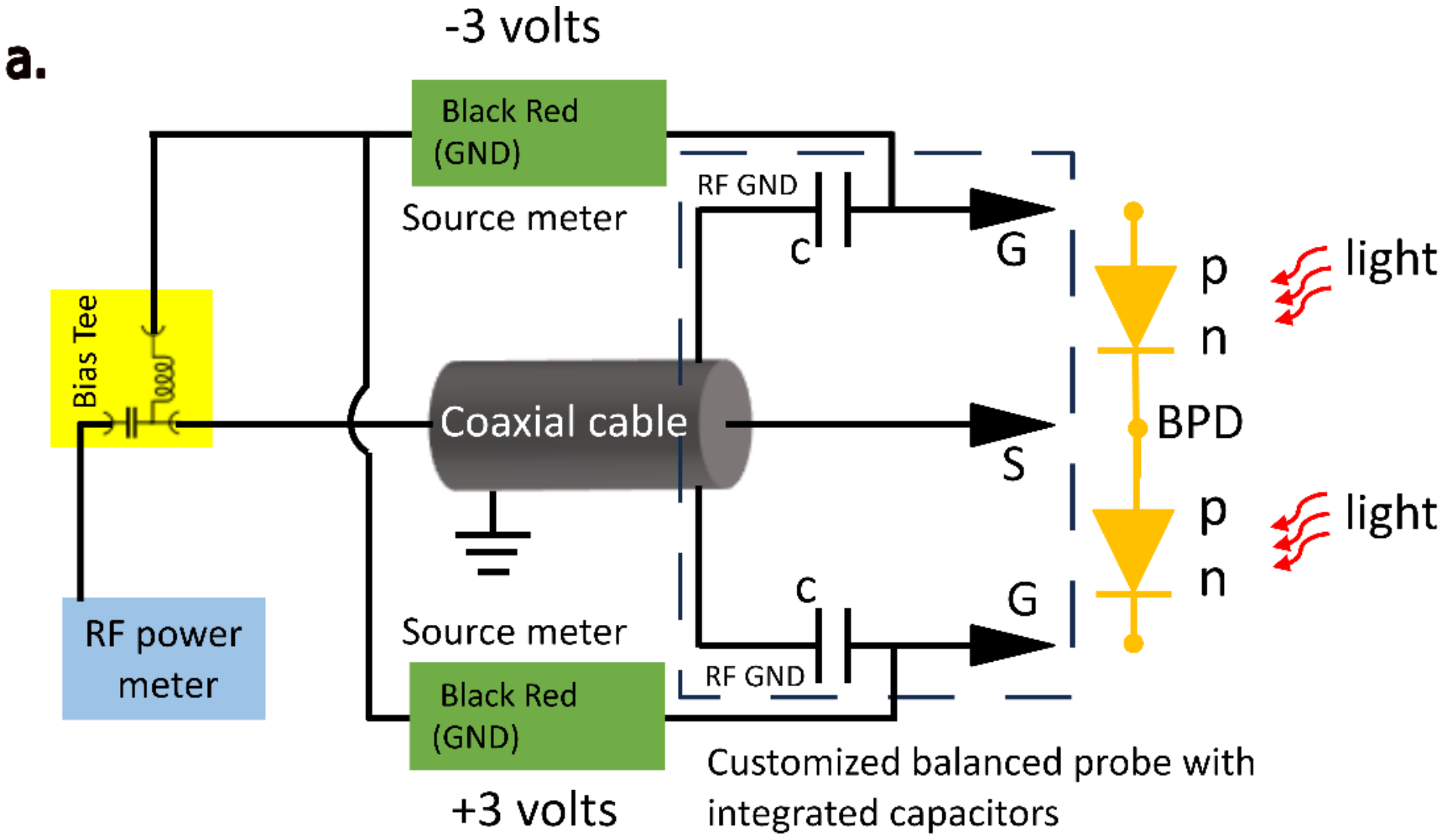
a.
-3 volts
Black Red
(GND)
Source meter
RF GND
c
G
p
n
light
Bias Tee
Coaxial cable
BPD
S
p
light
n
c
G
Source meter
RF GND
RF power
meter
Black Red
(GND)
+3 volts
Customized balanced probe with
integrated capacitors

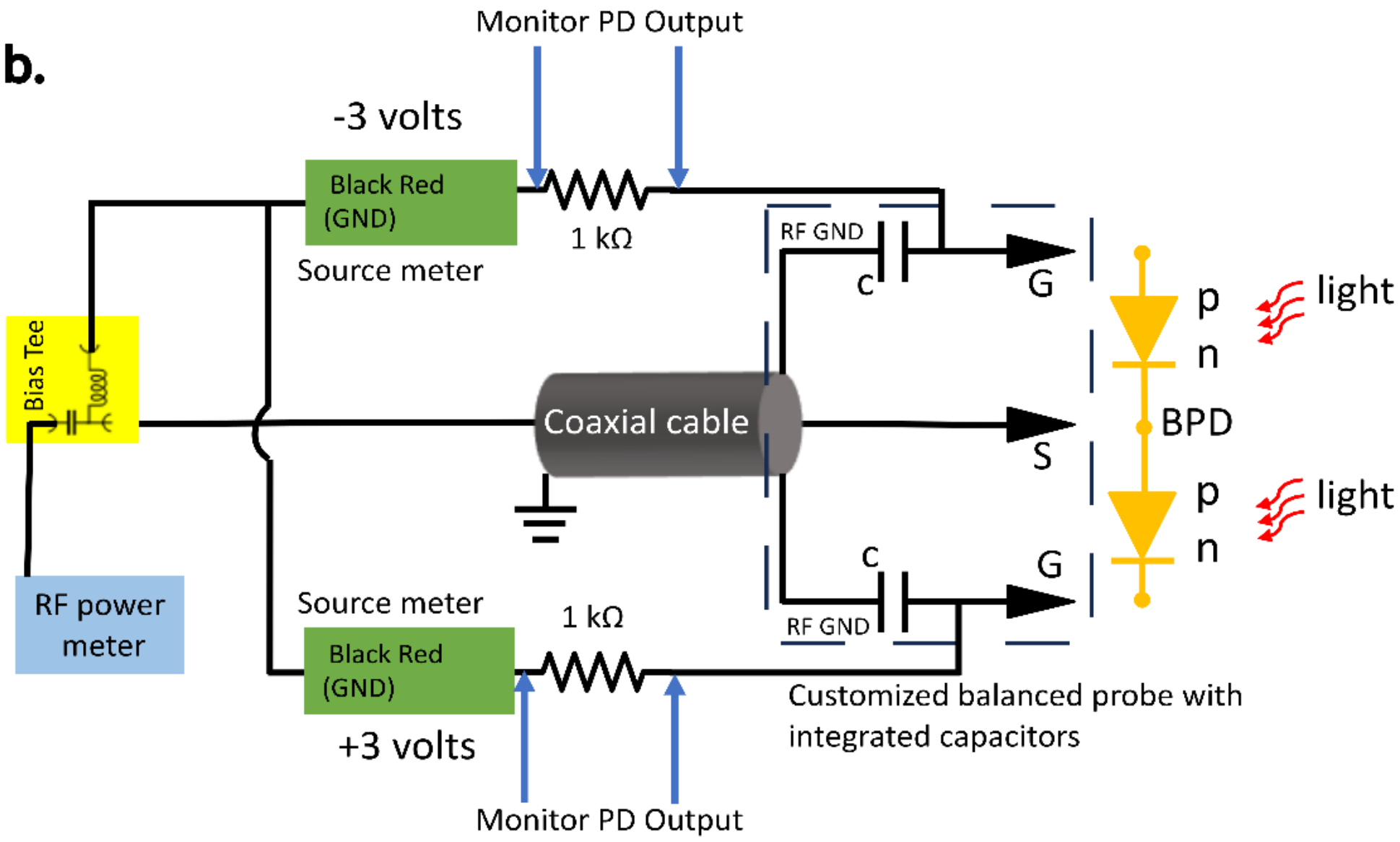
b.
Monitor PD Output
-3 volts
Black Red
(GND)
1 kΩ
Source meter
RF GND
c
G
p
light
n
Bias Tee
Coaxial cable
BPD
S
p
light
n
c
G
RF power
meter
Source meter
1 kΩ
RF GND
Black Red
(GND)
+3 volts
Customized balanced probe with
integrated capacitors
Monitor PD Output

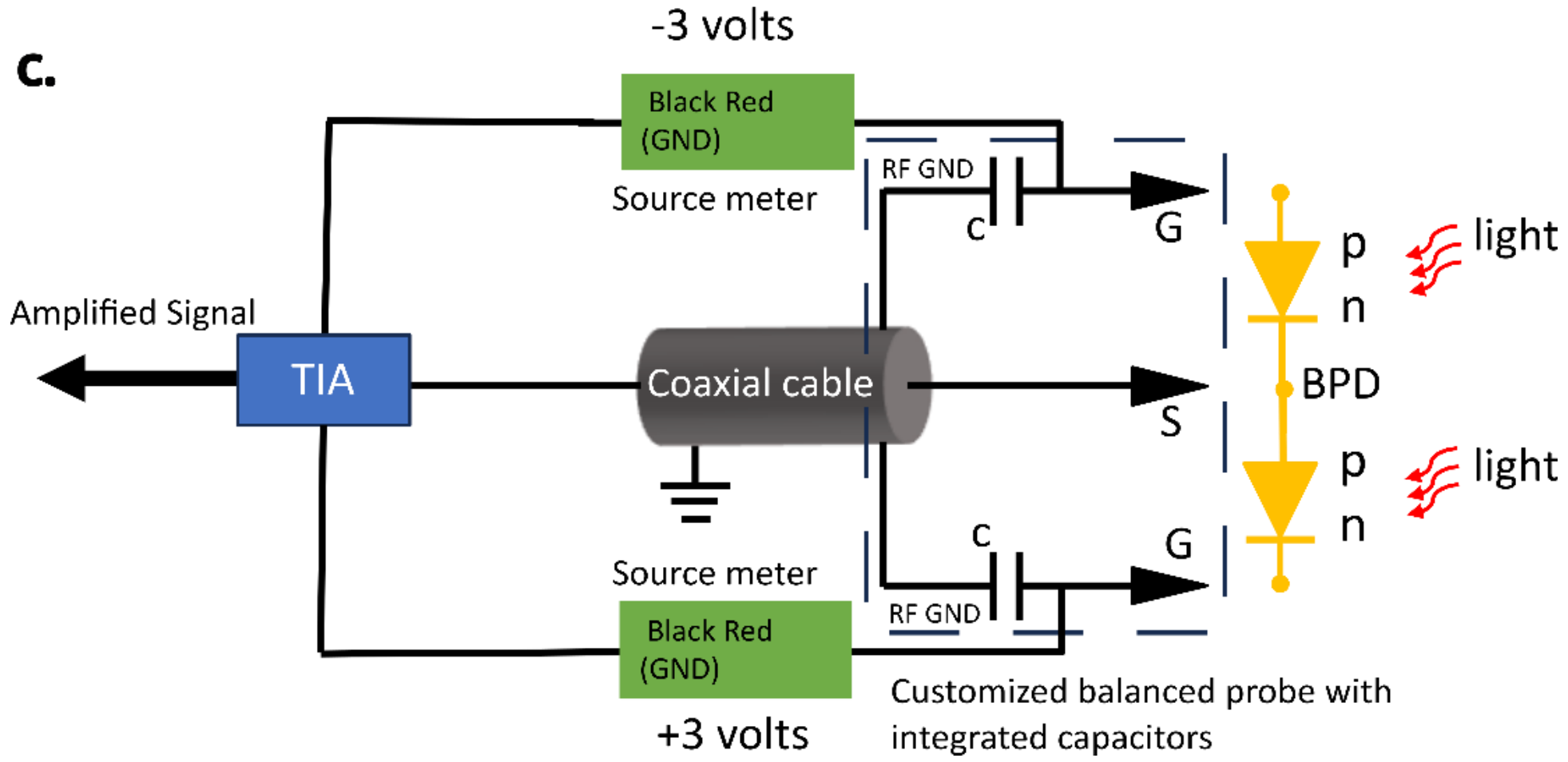
c.
-3 volts
Black Red
(GND)
Source meter
RF GND
c
G
p
light
n
Amplified Signal
TIA
Coaxial cable
BPD
S
p
light
n
c
G
Source meter
RF GND
Black Red
(GND)
+3 volts
Customized balanced probe with
integrated capacitors

***Supplemental Fig. 2. Electrical Setups used for accessing the Heterogeneously Integrated BPD and MZI.*** *(a) Electrical setup used to obtain complete, unamplified AC signal. (b) Electrical setup used to obtain detected signals on each individual PD. (c) Electrical setup used for laser stabilization and frequency noise measurement with the TIA.* ***GND****: Ground.* ***BPD****: Balanced Photodetector.* ***TIA****: Transimpedance Amplifier.*

**Section III. Transimpedance Amplifier Details and Bandwidth**

A custom made TIA is fabricated on a PCB with a basic schematic shown in Supplemental Fig. 3. The op-amp used is a Texas Instruments OPA846 while the gain offered by the resistor is $10^4$. The circuit had power rails at a ±5 V bias with blocking capacitors. The output had a 50 Ohm impedance.

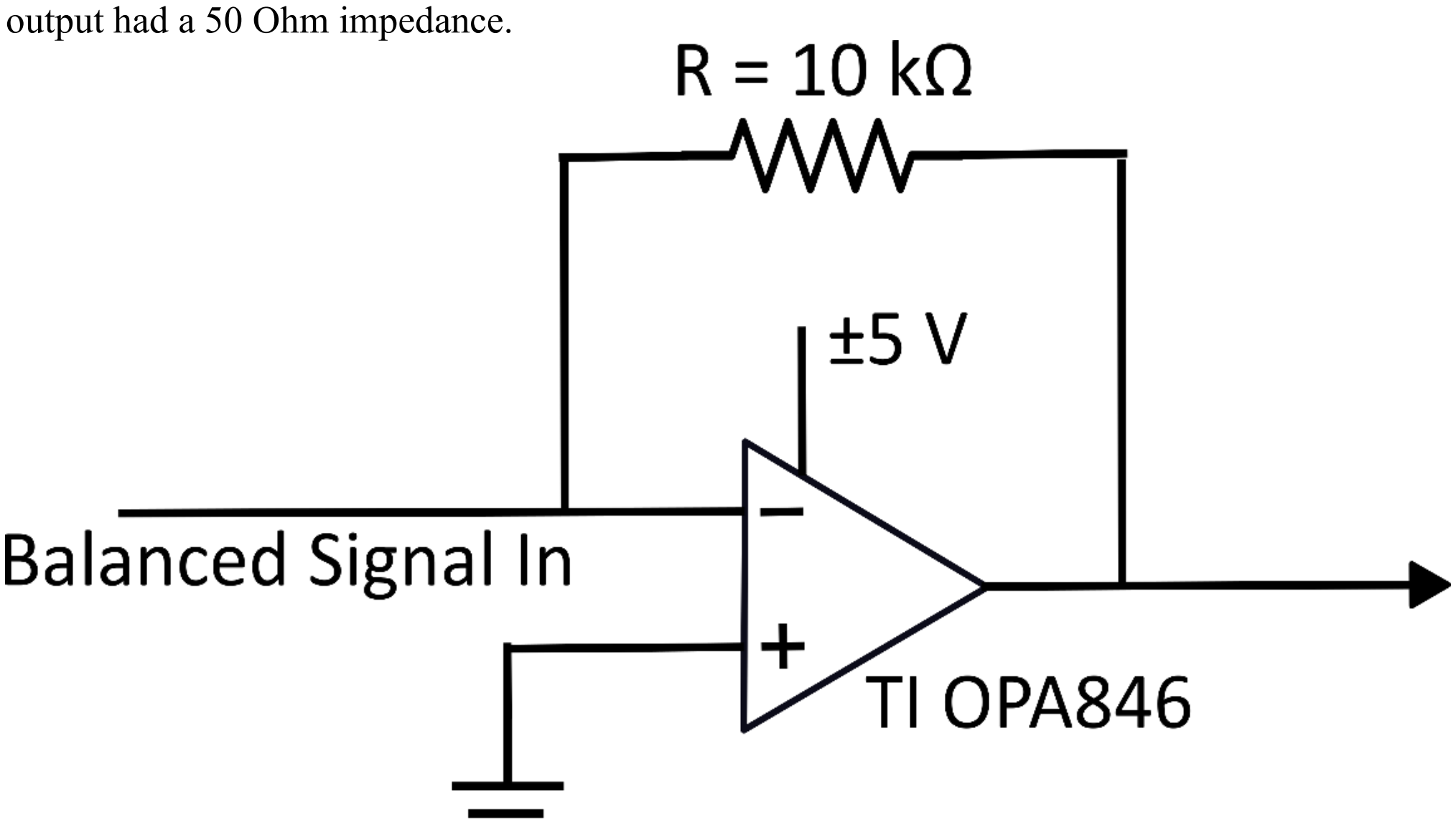

***Supplemental Fig. 3. Schematic of the TIA.*** *A circuit schematic of the TIA used for laser stabilization and frequency noise measurements collecting the output of the BPD. The overall gain of the circuit is 10000.* ***TIA****: Transimpedance Amplifier.*

An analysis of the combined bandwidth of the TIA with the heterogeneously integrated BPD and MZI follows closely to the setup seen in Fig. 6a. However, here the TIA is added before the PNA and no EDFA is used. The listed bandwidth of OPA846 is 20 MHz [1]. The frequency response of this system can be found in Supplemental Fig. 4.

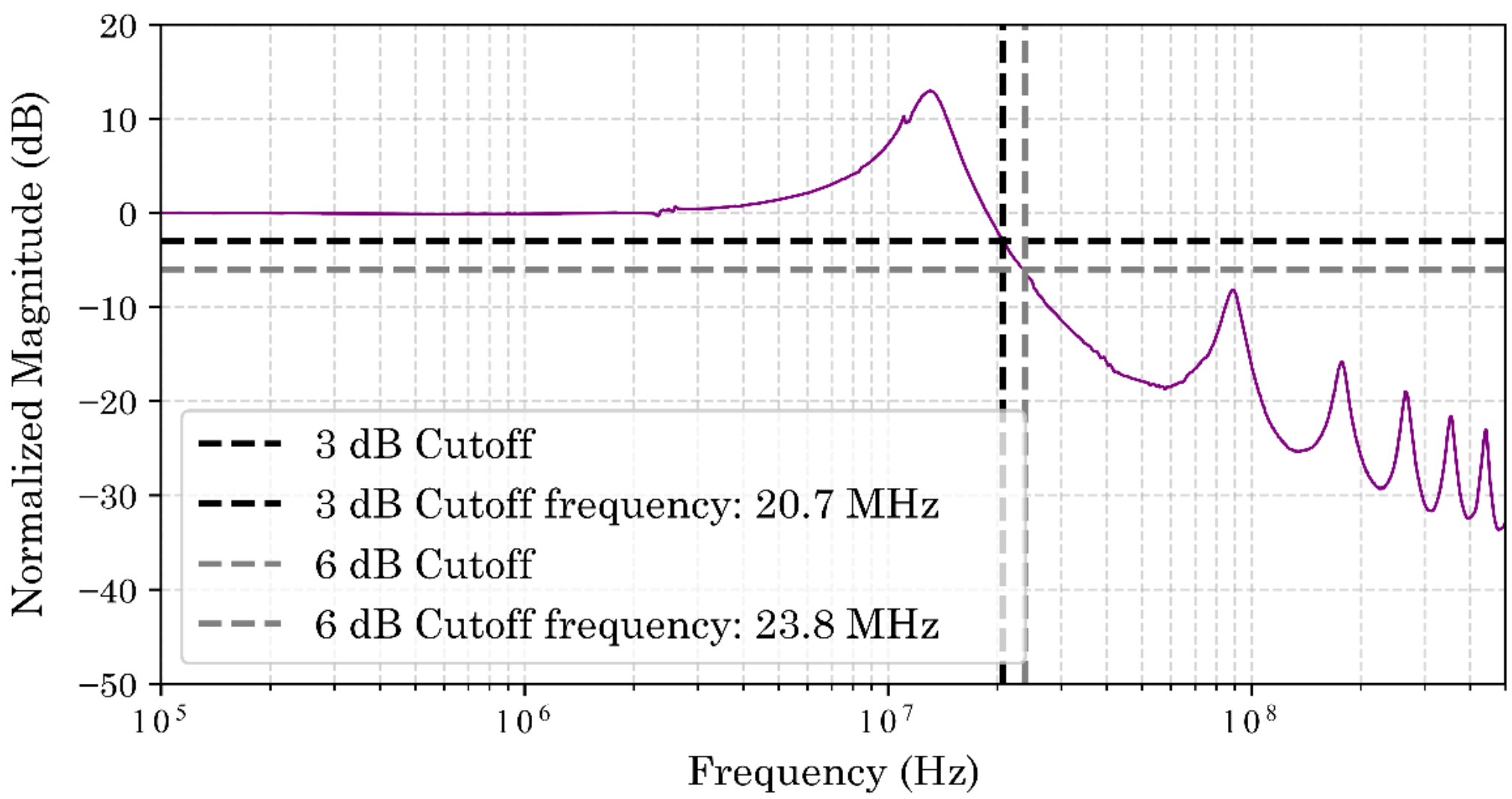

***Supplemental Fig. 4. Total Bandwidth of the Heterogeneously Integrated BPD and MZI with the TIA.*** *A lower 3 dB cutoff frequency is seen here with the TIA in comparison to without (Fig. 6b). Peaks due to ripple effects.* ***TIA****: Transimpedance Amplifier.* ***MZI****: Mach-Zehnder Interferometer.* ***BPD****: Balanced Photodetector.*

**Section IV. Analysis of MZI Frequency Response on Bandwidth**

We note that the amplitude modulation (AM) transfer function of an MZI is given by the following [2]:

$$H(f) = \cos(\pi f \tau)\ e^{-j\pi f \tau} \tag{S1}$$

Where f is the modulation frequency and τ is $1/_{FSR}$ or the time delay of the delay arm. Accounting for the extinction ratio, the magnitude response is given by the following:

$$\mathrm{S(f)} = 10\log_{10}\big(\epsilon|\mathrm{H(f)}| + (1-\epsilon)\big) \tag{S2}$$

Where ε is the measured extinction ratio of the MZI in linear units.